\documentclass{article}
\usepackage{spconf}
\usepackage{amsmath,amssymb,epsfig,textcomp}
\usepackage{graphics,bm,enumerate,multirow,cite,subfigure,setspace,array}
\usepackage[english]{babel}
\usepackage{amssymb,amsmath,epsfig}
\usepackage{latexsym}
\usepackage{graphicx,threeparttable,float,balance,array,bm,caption}
\usepackage{cite}
\usepackage{hyperref}

\title{Target Speaker Extraction for Overlapped Multi-Talker Speaker Verification}

\makeatletter
\def\name#1{\gdef\@name{#1\\}
}
\makeatother \name{{\em Wei Rao$^{1,2}$ \thanks{This work is submitted to INTERSPEECH 2019.}, Chenglin Xu$^{2,3}$, Eng Siong Chng$^{2,3}$, Haizhou Li$^{1,2}$}}

\address{
$^{1}$Department of Electrical and Computer Engineering, National University of Singapore, Singapore\\
$^{2}$Temasek Laboratories, Nanyang Technological University, Singapore \\
$^{3}$School of Computer Science and Engineering, Nanyang Technological University, Singapore\\
}

\begin{document}
%
\maketitle
\begin{abstract}
The performance of speaker verification degrades significantly when the test speech is corrupted by interference speakers. Speaker diarization does well to separate speakers if the speakers are temporally overlapped. However, if multi-talkers speak at the same time, we need the technique to separate the speech in the spectral domain. This paper proposes an overlapped multi-talker speaker verification framework by using target speaker extraction methods. Specifically, given the target speaker information, the target speaker's speech is firstly extracted from the overlapped multi-talker speech by a target speaker extraction module. Then, the extracted speech is passed to the speaker verification system. Experimental results show that the proposed approach significantly improves the performance of overlapped multi-talker speaker verification and achieves 65.7\% relative EER reduction.

\end{abstract}
\vspace{-6pt}
\section{Introduction}
\label{sec:intro}
The performance of speaker verification is significantly degraded when the speech contains background noise and/or is corrupted by interference speakers. Speaker diarization is usually applied on the non-overlapped multi-talker speech by speaker segmentation and clustering~\cite{anguera2012speaker}. It still works well by detecting and excluding the overlapped speech when the multi-talker speech is slightly overlapped \cite{charlet2013impact,yella2014overlapping}. However, such system fails when multi-talkers speak at the same time.

One possible solution is to separate the multi-talker speech into different speakers using a speech separation system, such as deep clustering \cite{hershey2016deep}, deep attractor network \cite{chen2017deep}, permutation invariant training \cite{kolbaek2017multitalker, xu2018single, xu2018shifted}, and so on. Although the performance of speech separation has been significantly improved by such approaches, the number of speakers has to be known in prior for these approaches. However, the number of speakers in the test speech is unknown in the real application of speaker verification.  


To address the limitation that the number of speakers has to be know in prior, this paper proposes an overlapped multi-talker speaker verification framework by using the target speaker extraction. Given the target speaker information, the target speaker's speech is extracted from the overlapped multi-talker speech by a target speaker extraction module. Although the target speaker extraction system needs the target speaker information, it will not be a limitation. Because the target speaker information is provided as enroll speech in speaker verification system. Then, the extracted speech is passed to the speaker verification system (i.e. i-vector/PLDA speaker verification system \cite{Dehak&Kenny2011, Kenny10, Prince07, Garcia11} in this work) to verify whether the extracted speech is belonging to the target speaker. 


Moreover, this paper also compares the effectiveness of two target speaker extraction networks for the overlapped multi-talker speaker verification. They are SBF-MTSAL \cite{Xu&Rao2019spkext} and SBF-MTSAL-Concat \cite{Xu&Rao2019spkext}. Experimental results demonstrate that the proposed method in this paper significantly improve the performance of speaker verification on overlapped multi-talker speech. In addition, SBF-MTSAL-Concat outperforms  SBF-MTSAL in the overlapped multi-talker speaker verification.

The remainder of the paper is organized as follows.  Section~\ref{sec:tse_sv} introduces our proposed overlapped multi-talker speaker verification framework in this paper. Section~\ref{sec:exp} and Section~\ref{sec:exp_res} report the experimental setup and results. Then, the conclusions and future works are presented in Section~\ref{sec:con}. 


\vspace{-6pt}
\section{Multi-Talker Speaker Verification with Speaker Extraction}
\label{sec:tse_sv}
\begin{figure}[t]
  \centering
  \centerline{\includegraphics[width=\linewidth]{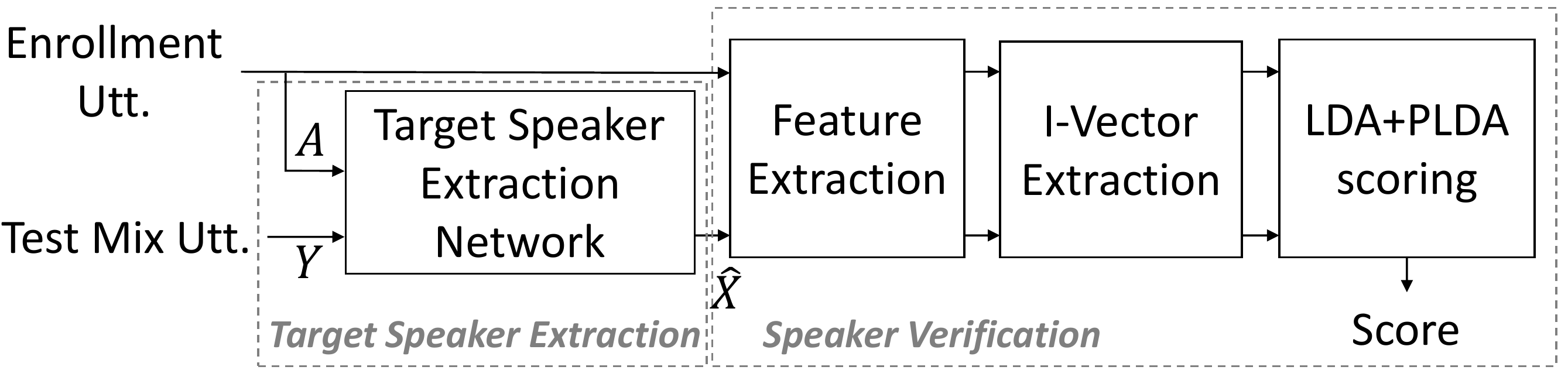}}
  \caption{The flow chart of overlapped multi-talker speaker verification system with target speaker extraction. ``$Y$'' represents the mixture speech. ``$A$'' represents the enrollment speech. ``$\hat{X}$'' represents the extracted target speaker's speech from $Y$. The target speaker extraction network takes the enrollment utterance $A$ of the target speaker as the auxiliary information to extract the speech component $\hat{X}$ from $Y$ that belongs to the target speaker.}
  \label{fig:overall_frm}
  \vspace{-8pt}
\end{figure}

Since the target speaker information will be given in speaker verification, target speaker extraction is a good option to address the overlapped multi-talker speaker verification problem. Fig.~\ref{fig:overall_frm} illustrates the framework of the proposed overlapped multi-talker speaker verification system with target speaker extraction. The framework consists of a target speaker extraction module and a speaker verification system. Specifically, given a trial, the enrollment utterance $A$ and the overlapped multi-talker test utterance $Y$ are fed into the target speaker extraction network. Then, the target speaker's speech $\hat{X}$ is extracted from $Y$. The extracted speech $\hat{X}$ and enrollment speech $A$ are used as inputs of the standard i-vector/PLDA speaker verification system \cite{Dehak&Kenny2011,Kenny10,Prince07,Garcia11} to verify whether the extracted speech is belonging to the target speaker.


In this paper, we compare two target speaker extraction methods in the multi-talker speaker verification framework: (1) SpeakerBeam front-end with magnitude and temporal spectrum approximation loss (SBF-MTSAL) \cite{Xu&Rao2019spkext} and (2) SBF-MTSAL with concatenation framework (SBF-MTSAL-Concat) \cite{Xu&Rao2019spkext}. Both of these two methods are  the extended methods of SpeakerBeam front-end (SBF) \cite{delcroix2018single, delcroix2016context}.


\begin{figure}[t]
  \centering
  \centerline{\includegraphics[width=7.0cm]{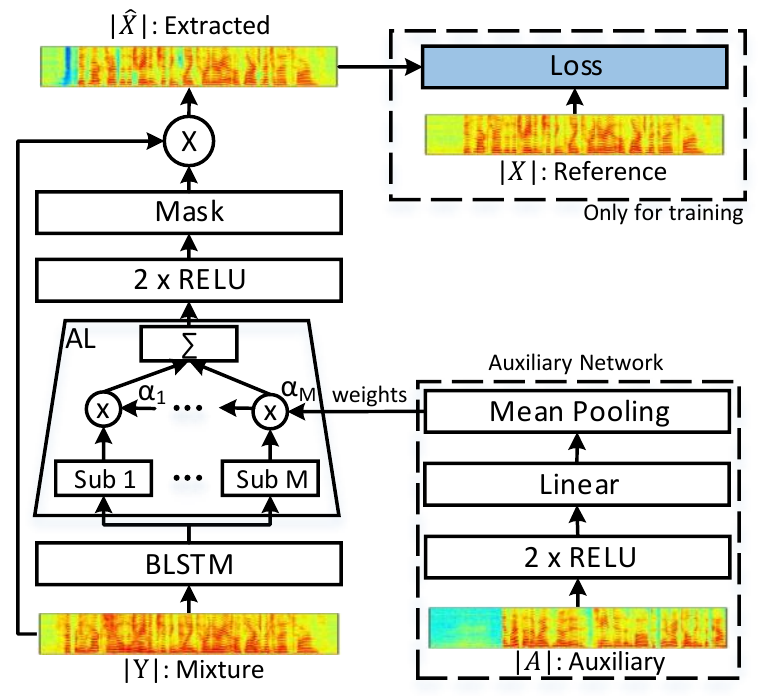}}
  \caption{The architecture of SBF-MTSAL. ``AL'' in the trapezium box represents the adaptation layer. ``Sub'' represents the sub-layer. ``$\alpha$'' represents the weight obtained from the auxiliary network. ``$M$'' represents the number of sub-layers. ``$|Y|:$ Mixture'' represents the magnitude of the mixture speech. ``$|\hat{X}|:$ Extracted'' represents the output magnitude of the extracted target speaker's speech. ``$|X|:$ Reference'' represents the magnitude of clean speech, which is used to simulate the mixture. ``$|A|:$ Auxiliary'' represents the magnitude of the auxiliary speech. During the evaluation, the upper right dotted box is not necessary.}
  \label{fig:speakerbeam_frm}
\vspace{-8pt}
\end{figure}

\begin{figure}[t]
  \centering
  \centerline{\includegraphics[width=6.5cm]{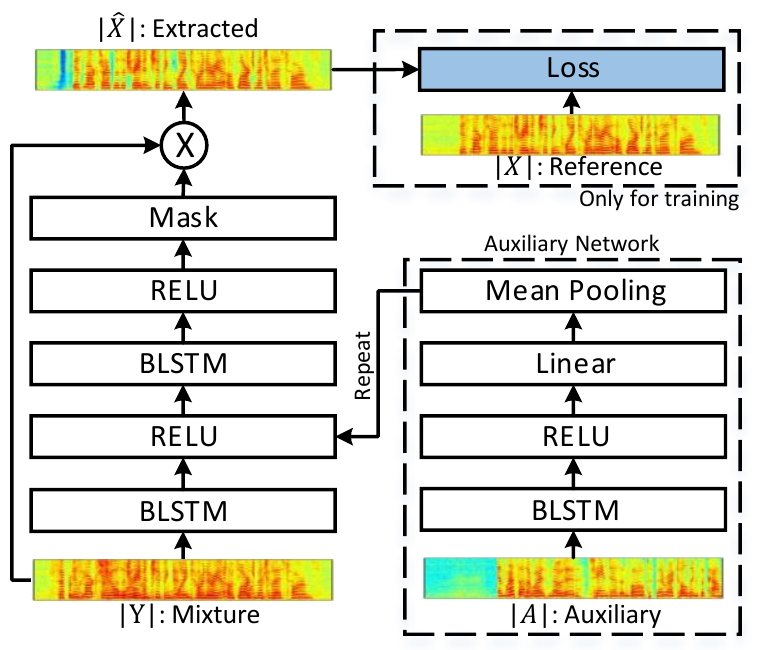}}
    \caption{The architecture of SBF-MTSAL-Concat. The meaning of ``$|Y|, |\hat{X}|, |X|, |A|$'' can be referred to the caption of Fig.~\ref{fig:speakerbeam_frm}. During the evaluation, the upper right dotted box is not necessary.}
    \label{fig:josen_frm}
\vspace{-8pt}
\end{figure}

\vspace{-10pt}
\subsection{SBF-MTSAL}
Fig.~\ref{fig:speakerbeam_frm} shows the architecture of SBF-MTSAL \cite{Xu&Rao2019spkext}. The SBF-MTSAL approach uses an auxiliary network to learn adaptation weights from the target speaker's voice, which is different from the utterance of the target speaker in the mixture. The adaptation weights contain speaker characteristics and are used to weight the sub-layers in the adaptation layer of the mask estimation network with a CADNN structure. Instead of computing objective loss between ideal binary mask and estimated mask in the original work \cite{delcroix2018single}, SBF-MTSAL computes a magnitude and temporal spectrum approximation loss to estimate a phase sensitive mask \cite{erdogan2015phase} due to its better performance. The magnitude and its dynamic information (i.e., delta and acceleration) are used in calculating the objective loss for temporal continuity. 

\vspace{-10pt}
\subsection{SBF-MTSAL-Concat}
Fig.~\ref{fig:josen_frm} illustrates the architecture of SBF-MTSAL-Concat method \cite{Xu&Rao2019spkext}. The auxiliary network learns speaker embedding from a different utterance of target speaker, which contains speaker characteristics. Then, the speaker embedding is repeated concatenated with the activation of a BLSTM in the mask estimation network. The concatenated representations containing the mixture and target speaker information are used to estimate a phase sensitive mask with the same loss function as in SBF-MTSAL method. 

\vspace{-6pt}
\section{Experimental Setup}
\label{sec:exp}
\vspace{-6pt}

\subsection{Speech Data}\label{sec:exp_data}
\vspace{-6pt}
The two-speaker mixed dataset\footnote{The database simulation code is available at: \url{https://github.com/xuchenglin28/speaker_extraction}} used to train the target speaker extraction network was simulated at a sampling rate of 8kHz based on the WSJ0 corpus \cite{garofolo1993csr}. In the simulation of two-speaker mixture, the first selected speaker was chosen as target speaker, the other one was interference speaker. The utterance of the target speaker from the original WSJ0 corpus was used as reference speech. Another utterance of this target speaker, which was different from the reference speech, was randomly selected to be used as input to the auxiliary network to obtain target speaker information.

The simulated dataset was divided into training set ($20,000$ utterances), development set ($5,000$ utterances), and test set ($3,000$ utterances). Specifically, the utterances from $50$ male and $51$ female speakers in the WSJ0 ``si\_tr\_s'' set were randomly selected to generate the training and development set. The SNR of each mixture was randomly selected between 0dB and 5dB. Similarly, the test set was created by randomly mixing the utterances from $10$ male and $8$ female speakers in the WSJ0 ``si\_dt\_05'' and ``si\_et\_05'' sets. Since the speakers in the test set were different from the training and development sets, the test set was used to evaluate the speaker verification performance.
\vspace{-6pt}

\subsection{Target Speaker Extraction Network Setup}\label{sec:exp_tse}
\vspace{-6pt}

A short-time Fourier transform (STFT) was used with a window length of $32$ms and a shift of $16$ms to obtain the magnitude features from both of the input mixture for mask estimation network and input target speech for auxiliary network. The normalized square root hamming window was applied.

The learning rate started from $0.0005$ and scaled down by $0.7$ when the training loss increased on the development set. The minibatch size was set to $16$. The network was trained with minimum $30$ epochs and stopped when the relative loss reduction was lower than $0.01$. The Adam algorithm \cite{kingma2014adam} was used to optimize the network.

The aforementioned magnitude extraction configuration and network training scheme were kept same in both SBF-MTSAL and SBF-MTSAL-Concat methods. The extracted magnitude were reconstructed into time-domain signal with phase of the mixture. Then the time-domain signal was used as input to the speaker verification system.


For SBF-MTSAL, the auxiliary network was composed of $2$ feed-forward relu hidden layers with $512$ hidden nodes and a linear layer with $30$ hidden nodes. The adaptation weights were obtained by averaging these $30$ dimensional outputs over all the frames. The mask estimation network used a BLSTM with $512$ cells in each forward and backward direction. The following adaptation layer had $30$ sub-layers. Each sub-layer had $512$ nodes with $1024$ dimensional inputs from the outputs of the previous BLSTM. The $30$ dimensional weights from the auxiliary network were used to weight these sub-layers, respectively. Then the activation of the adaptation layer was summed over all the sub-layers. Another $2$ feed-forward relu hidden layers with $512$ nodes were appended. The mask layer had $129$ nodes to predict the mask for the target speaker.

\begin{table*}[t]
    \centering
    \caption{Performance of SV system without and with target speaker extraction. ``Training'' represents the type of training data for SV system. ``Eval.'' represents the type of evaluation test data for SV system. ``TSE'' represents whether perform the target speaker extraction for SV. If the option of this column is No, it means that no speaker extraction is applied in the SV. If the option is SBF-MTSAL/SBF-MTSAL-Concat, it means that SBF-MTSAL/SBF-MTSAL-Concat speaker extraction is used for SV. ``Clean'' represents the speech from one speaker. ``Mixture'' means the overlapped multi-talker's speech. ``Ext'' represents the extracted target speaker's speech from the mixture by target speaker extraction network. ``Clean+Ext'' means pooling the clean speeches and extracted target speaker's speeches to train the SV system. ``Baseline'' represents the baseline performance of overlapped multi-talker speaker verification system. ``Upper Bound'' represents the upper bound performance of target speaker extraction for overlapped multi-talker speaker verification. ``DCF08'' represents the minimum detection cost with $P_{Target}=0.01$. ``DCF10'' represents the minimum detection cost with $P_{Target}=0.001$. The details of experimental setup can be referred to section~\ref{sec:exp_sv}.}
    \vspace{-6pt}
    \begin{tabular}{|l|l|l|l|c|c|c|} \hline
      System No. & Training & Eval. & TSE & EER $(\%)$ & DCF08 & DCF10 \\ \hline\hline
      1 (Baseline) & Clean & Mixture & No & 22.67 & 0.867 & 0.915 \\
      2 & Clean+Ext & Mixture & No & 21.67 & 0.850 & 0.898 \\ \hline\hline
      3 & Clean & Mixture & SBF-MTSAL & 11.17 & 0.760 & 0.844 \\ 
      4 & Clean & Mixture & SBF-MTSAL-Concat & 10.40 & 0.736 & 0.813 \\ 
      5 & Clean+Ext & Mixture & SBF-MTSAL-Concat & \textbf{7.77} & \textbf{0.631} & \textbf{0.747} \\ \hline\hline
      6 (Upper Bound) & Clean & Clean & No & 3.33 & 0.357 & 0.454 \\ 
      7 & Clean+Ext & Clean & No & 3.07 & 0.377 & 0.524 \\ \hline
    \end{tabular}
    \label{tab:SV_I}
    \vspace{-12pt}
\end{table*}


For SBF-MTSAL-Concat, the auxiliary had a BLSTM with $256$ cells in each forward and backward direction, a feed-forward relu hidden layer with $256$ nodes and a linear layer with $30$ nodes. The output of the linear layer was averaged over all frames to obtain a $30$ dimensional speaker embedding containing target speaker characteristics. The speaker embedding was repeatedly concatenated with the activation of the BLSTM layer in the mask estimation network. The BLSTM had $512$ cells in each forward and backward direction. Then the concatenated outputs were fed to a feed-forward relu hidden layer, a BLSTM layer and another feed-foreard relu hidden layer. The BLSTM had $512$ cells and the relu layers had $512$ nodes. The mask layer had $129$ nodes.
\vspace{-6pt}

\subsection{Speaker Verification (SV) System}\label{sec:exp_sv}
\vspace{-6pt}

According to the test set of simulated dataset, we generated 3000 target trials and 48,000 non-target trials for the SV evaluation. In the evaluation trials, each enrollment utterance contained a contiguous speech segment from a single speaker and test utterance contained the overlapped speech from multiple speakers. We called this evaluation set as \textit{mixture evaluation set}. Moreover, to show the upper bound of target speaker extraction on SV, we also generated another evaluation set with 51,000 trials from WSJ0 corpus according to the information of mixture set. This set was called as \textit{clean evaluation set}\footnote{The SV evaluation trials and keys for clean and mixture evaluation sets are available at: \url{https://github.com/xuchenglin28/speaker_extraction}}.


We selected 8,769 utterances from 101 speakers in WSJ0 corpus which were used to generate the training set of simulated database for training UBM, total variability matrix, LDA, and PLDA models. This set was named as \textit{clean training set}. Because this paper directly used the extracted target speaker's speech for SV, it would cause the mismatch between extracted speech and clean speech. To solve this mismatched problem, we pooled 5,000 extracted speech from the development set of the simulated 2-speaker mixed dataset and clean training set to train speaker verification system. We called this training set as \textit{clean+ext set}. Section~\ref{sec:exp_res} will show the performance by using different training and evaluation set.

The features of SV system were based on 19 MFCCs together with energy plus their 1st- and 2nd-derivatives extracted from the speech regions, followed by cepstral mean normalization \cite{Atal74} with a window size of 3 seconds. A 60-dimensional acoustic vector is extracted every 10ms, using a Hamming windowing of 25ms. An energy based voice activity detection method is used to remove silence frames. The system was based on a gender-independent UBM with 512 mixtures. The training set described in the previous paragraph was used for estimating the UBM and total variability matrix with 400 total factors. The same data set was used for estimating the LDA and Gaussian PLDA models with 150 latent variables.

\vspace{-6pt}
\section{Experimental Results}
\label{sec:exp_res}
\vspace{-6pt}

To investigate the effect of overlapped test speech on speaker verification, we perform the speaker verification experiments on both mixture and clean evaluation set described in section~\ref{sec:exp_sv}. System 1 of table~\ref{tab:SV_I} is the baseline system of SV with clean training data on mixture test set. System 6 of table \ref{tab:SV_I} shows the upper bound performance (also called as ideal performance) of target speaker extraction for overlapped multi-talker SV. Performance comparison between system 1 and system 6 of table~\ref{tab:SV_I} shows that the performance speaker verification system seriously degrades when the test speech is the fully overlapped multi-talker speech.   

Table~\ref{tab:SV_I} also presents the performance of speaker verification system on the evaluation set without and with target speaker extraction. System 1 of table~\ref{tab:SV_I} is the baseline results of overlapped multi-talker speaker verification. System 3 to 5 of table~\ref{tab:SV_I} show the performance of speaker verification after using target speaker extraction. Results of system 1 to 5 demonstrate that applying target speaker extraction significantly improve the performance of overlapped multi-talker speaker verification. Specifically, System 5 of Table~\ref{tab:SV_I} can obtain around $65.7\%, 27.2\%, 18.4\%$ relative reduction over the baseline (system 1) on EER, DCF08, and DCF10, respectively.

This paper compares two target speaker extraction methods for overlapped multi-talker speaker verification: (1) SBF-MTSAL and (2) SBF-MTSAL-Concat. Performance comparison between system 3 and 4 of table~\ref{tab:SV_I} shows that SBF-MTSAL-Concat outperforms SBF-MTSAL on both EER and DCFs. 

To alleviate the effect of the mismatch between extracted speech and clean speech for SV, we combine the clean training set and extracted speech data (Clean+Ext) to train speaker verification system. Both clean and mixture test sets are used to evaluation this speaker verification system. Additionally, because SBF-MTSAL-Concat achieves the better performance, we only apply this experiment on SBF-MTSAL-Concat. System 2, 5, and 7 in table~\ref{tab:SV_I} show the performance of SV system with clean+ext training data. System 2 of table~\ref{tab:SV_I} proves that most of improvement on the overlapped multi-talker speaker verification is done by target speaker extraction methods by comparing with system 1 and 3 to 5. Comparison between system 6 and 7 in table~\ref{tab:SV_I} shows that Clean+Ext training set will improve the performance of speaker verification on the clean test set in EER, but degrade the DCFs. And performance comparison between system 4 and 5 in table~\ref{tab:SV_I} demonstrates that Clean+Ext training set could advance the speaker verification performance with SBF-MTSAL-Concat on the mixture test set.

\vspace{-12pt}
\section{Conclusions and Future Works}
\label{sec:con}
\vspace{-6pt}

This paper applies the target speaker extraction to improve the performance of overlapped multi-talker speaker verification. Experimental results show that the proposed method could significantly improve the performance of overlapped multi-talker speaker verification. This paper also compares the performance of SBF-MTSAL and SBF-MTSAL-Concat on overlapped multi-talker speaker verification and finds that SBF-MTSAL-Concat achieves better performance than SBF-MTSAL. 


This paper mainly focuses on the fully overlapped test speech from multiple speakers.
In the future, we will investigate the effectiveness of proposed method when the enrollment is also a multi-talker speech, apply the proposed method on the public available speaker verification database, and explore the joint training of target speaker extraction and speaker verification.  

\vspace{-6pt}

\bibliographystyle{IEEEbib}
\bibliography{ref_db1,mypapers}

\begin{thebibliography}{10}

\bibitem{anguera2012speaker}
Xavier Anguera, Simon Bozonnet, Nicholas Evans, Corinne Fredouille, Gerald
  Friedland, and Oriol Vinyals,
\newblock ``Speaker diarization: A review of recent research,''
\newblock {\em IEEE Transactions on Audio, Speech, and Language Processing},
  vol. 20, no. 2, pp. 356--370, 2012.

\bibitem{charlet2013impact}
Delphine Charlet, Claude Barras, and Jean-Sylvain Li{\'e}nard,
\newblock ``Impact of overlapping speech detection on speaker diarization for
  broadcast news and debates,''
\newblock in {\em Proc. of ICASSP}. IEEE, 2013, pp. 7707--7711.

\bibitem{yella2014overlapping}
Sree~Harsha Yella and Herv{\'e} Bourlard,
\newblock ``Overlapping speech detection using long-term conversational
  features for speaker diarization in meeting room conversations,''
\newblock {\em IEEE/ACM Transactions on Audio, Speech, and Language
  Processing}, vol. 22, no. 12, pp. 1688--1700, 2014.

\bibitem{hershey2016deep}
John~R Hershey, Zhuo Chen, Jonathan Le~Roux, and Shinji Watanabe,
\newblock ``Deep clustering: Discriminative embeddings for segmentation and
  separation,''
\newblock in {\em Proceedings of ICASSP}. IEEE, 2016, pp. 31--35.

\bibitem{chen2017deep}
Zhuo Chen, Yi~Luo, and Nima Mesgarani,
\newblock ``Deep attractor network for single-microphone speaker separation,''
\newblock in {\em Proceedings of ICASSP}. IEEE, 2017, pp. 246--250.

\bibitem{kolbaek2017multitalker}
Morten Kolb{\ae}k, Dong Yu, Zheng-Hua Tan, and Jesper Jensen,
\newblock ``Multitalker speech separation with utterance-level permutation
  invariant training of deep recurrent neural networks,''
\newblock {\em IEEE/ACM Transactions on Audio, Speech, and Language
  Processing}, vol. 25, no. 10, pp. 1901--1913, 2017.

\bibitem{xu2018single}
Chenglin Xu, Wei Rao, Xiong Xiao, Eng~Siong Chng, and Haizhou Li,
\newblock ``Single channel speech separation with constrained utterance level
  permutation invariant training using grid lstm,''
\newblock in {\em Proceedings of ICASSP}. IEEE, 2018.

\bibitem{xu2018shifted}
Chenglin Xu, Wei Rao, Eng~Siong Chng, and Haizhou Li,
\newblock ``A shifted delta coefficient objective for monaural speech
  separation using multi-task learning,''
\newblock in {\em Proceedings of Interspeech}, 2018, pp. 3479--3483.

\bibitem{Dehak&Kenny2011}
N.~Dehak, P.~Kenny, R.~Dehak, P.~Dumouchel, and P.~Ouellet,
\newblock ``Front-end factor analysis for speaker verification,''
\newblock {\em IEEE Trans. on Audio, Speech, and Language Processing}, vol. 19,
  no. 4, pp. 788--798, May 2011.

\bibitem{Kenny10}
P.~Kenny,
\newblock ``Bayesian speaker verification with heavy-tailed priors,''
\newblock in {\em Proc. of Odyssey: Speaker and Language Recognition Workshop},
  Brno, Czech Republic, Jun. 2010.

\bibitem{Prince07}
S.J.D. Prince and J.H. Elder,
\newblock ``Probabilistic linear discriminant analysis for inferences about
  identity,''
\newblock in {\em Proc. of 11th International Conference on Computer Vision},
  Rio de Janeiro, Brazil, Oct. 2007, pp. 1--8.

\bibitem{Garcia11}
D.~Garcia-Romero and C.~Y. Espy-Wilson,
\newblock ``Analysis of i-vector length normalization in speaker recognition
  systems,''
\newblock in {\em Proc. of Interspeech 2011}, Florence, Italy, Aug. 2011, pp.
  249--252.

\bibitem{Xu&Rao2019spkext}
Chenglin Xu, Wei Rao, Eng~Siong Chng, and Haizhou Li,
\newblock ``Optimization of speaker extraction neural network with magnitude
  and temporal spectrum approximation loss,''
\newblock Accepted in ICASSP 2019.

\bibitem{delcroix2018single}
Marc Delcroix, Katerina Zmolikova, Keisuke Kinoshita, Atsunori Ogawa, and
  Tomohiro Nakatani,
\newblock ``Single channel target speaker extraction and recognition with
  speaker beam,''
\newblock in {\em Proceedings of ICASSP}. IEEE, 2018, pp. 5554--5558.

\bibitem{delcroix2016context}
Marc Delcroix, Keisuke Kinoshita, Chengzhu Yu, Atsunori Ogawa, Takuya Yoshioka,
  and Tomohiro Nakatani,
\newblock ``Context adaptive deep neural networks for fast acoustic model
  adaptation in noisy conditions,''
\newblock in {\em Proceedings of ICASSP}. IEEE, 2016, pp. 5270--5274.

\bibitem{erdogan2015phase}
Hakan Erdogan, John~R Hershey, Shinji Watanabe, and Jonathan Le~Roux,
\newblock ``Phase-sensitive and recognition-boosted speech separation using
  deep recurrent neural networks,''
\newblock in {\em Proceedings of ICASSP}. IEEE, 2015, pp. 708--712.

\bibitem{garofolo1993csr}
John Garofolo, D~Graff, D~Paul, and D~Pallett,
\newblock ``Csr-i (wsj0) complete ldc93s6a,''
\newblock {\em Web Download. Philadelphia: Linguistic Data Consortium}, 1993.

\bibitem{kingma2014adam}
Diederik Kingma and Jimmy Ba,
\newblock ``Adam: A method for stochastic optimization,''
\newblock {\em arXiv preprint arXiv:1412.6980}, 2014.

\bibitem{Atal74}
B.~S. Atal,
\newblock ``Effectiveness of linear prediction characteristics of the speech
  wave for automatic speaker identification and verification,''
\newblock {\em J. Acoust. Soc. Am.}, vol. 55, no. 6, pp. 1304--1312, Jun. 1974.

\end{thebibliography}

\end{document}